\documentclass[onecollarge,natbib]{svjour2}
\bibpunct{[}{]}{;}{n}{}{,} % to get "[numbered]" references from natbib
\smartqed  % flush right qed marks, e.g. at end of proof
\usepackage{graphicx}
\journalname{Archive of Applied Mechanics}
\begin{document}
\title{Nucleon Resonance Physics}%\thanks{Grants or other notes}

\author{Volker D. Burkert}

\institute{Volker D. Burkert \\
              Jefferson Laboratory \\
              12000 Jefferson Avenue, Newport News, Virginia 23606 \\
              Tel.: +1 757 269 7540\\
              Fax: +1 757 269 5800\\
              \email{burkert@jlab.org}  }

\date{Received: date / Accepted: date}
%The correct dates will be entered by the editor

\maketitle
\begin{abstract}
Recent results of meson 
photo-production at the existing electron machines with polarized real photon beams and 
the measurement of polarization observables of the final state baryons have provided 
high precision data that led to the discovery of new excited nucleon and $\Delta$ 
states using multi-channel partial wave analyses procedures. The internal structure of several 
prominent excited states has been revealed employing meson electroproduction processes. 
On the theoretical front, lattice QCD is now predicting the baryon spectrum with very similar 
characteristics as the constituent quark model, and continuum QCD, such as is represented 
in the Dyson-Schwinger Equations approach and in light front  relativistic quark models, 
describes the non-perturbative behavior of resonance excitations at photon virtuality of  
$Q^2 > 1.5GeV^2$. In this talk I discuss the need to continue a vigorous program of nucleon 
spectroscopy and the study of the internal structure of excited states as a way to reveal 
the effective degrees of freedom underlying the excited states and their dependence on 
the distance scale probed. 

\keywords{Nucleon resonances, meson electroproduction, helicity amplitudes, quark models}
\end{abstract}

\section{Introduction}
\label{intro}

The excited states of the nucleon have been studied experimentally 
since the 1950's~\cite{Anderson:1952nw}. They contributed to the discovery of the 
quark model in 1964 by 
Gell-Mann and Zweig~\cite{GellMann1964,Zweig1964}, and were critical for the discovery 
of "color" degrees of freedom as introduced by 
Greenberg~\cite{Greenberg:1964pe}. 
The quark structure of baryons resulted in the prediction of a wealth of excited states  
with underlying spin-flavor and orbital symmetry of $SU(6) \otimes O(3)$, and led to 
a broad experimental effort to search for these states. Most of 
the initially observed states were found with hadronic probes. However, of the many excited 
states predicted in the quark model, only a fraction have been observed to date.  
Search for the "missing" states and detailed studies of the resonance structure  
are now mostly carried out using electromagnetic probes and have been a major focus of  
hadron physics for the past decade \cite{Burkert:2004sk}. A broad  
experimental effort is currently underway with measurements 
of  exclusive meson photoproduction and electroproduction reactions, including 
many polarization observables. Precision data and the development of multi-channel 
partial wave analysis procedures have resulted in the discovery of several new excited states of the 
nucleon, which have been entered in the Review of Particle Physics~\cite{Agashe:2014kda}. 
 
A quantitative description of baryon spectroscopy and the structure of excited 
nucleons must eventually involve solving QCD for  
a complex strongly interacting multi-particle system. 
Recent advances in Lattice QCD led to predictions of the nucleon spectrum in QCD with 
dynamical quarks~\cite{Dudek:2012ag}, albeit with still large pion 
 masses of 396 MeV. Lattice prediction can therefore only be taken as indicative of the 
 quantum numbers of excited states and not of the masses of specific states. In parallel, 
 the development of dynamical coupled channel models is being pursued with new vigor. 
 The EBAC group at JLab has shown~\cite{Suzuki:2009nj} that dynamical effects can result 
 in significant mass shifts of the excited states. As a particularly striking result, a very large 
 shift was found for the Roper resonance pole mass to 1365 MeV downward from its bare 
 core mass of 1736 MeV. This result has clarified the longstanding puzzle of the incorrect 
 mass ordering of $N(1440){1\over 2}^+$ and $N(1535){1\over 2}^-$ resonances in the constituent 
 quark model. Developments on the phenomenological side go hand in hand with a 
 world-wide experimental effort to produce high precision data in many different channel 
 as a basis for a determination of the light-quark baryon 
 resonance spectrum. On the example of experimental results from CLAS, the strong impact 
 of precise meson photoproduction data is discussed.  
 Several reviews have recently been published on this and related 
 subjects~\cite{Klempt:2009pi,Tiator:2011pw,Aznauryan:2011qj,Aznauryan:2012ba,Crede:2013sze}. 

It is interesting to point out recent findings that relate the observed baryon spectrum of different 
quark flavors with the baryon densities in the freeze out temperature in heavy ion collisions,
which show evidence for missing baryons in the strangeness and the charm baryon 
sector~\cite{Bazavov:2014xya,Bazavov:2014yba}. These data hint that an improved baryon model
including further unobserved light quark baryons may resolve the current discrepancy between
lattice QCD results and the results obtained using a baryon model that includes only 
states listed by the PDG. A complete accounting of excited baryon states of all flavors seems 
essential for a quantitative description of the occurrence of baryons in the evolution of the 
microsecond old universe.     

Accounting for the complete excitation spectrum of the nucleon (protons and neutrons) 
and understanding the effective degrees of freedom is perhaps the most important and  
certainly the most challenging task of hadron physics. The experimental N* program currently 
focusses on the search for new excited states in the mass range from 2 GeV to 2.5 GeV 
using energy-tagged photon beams in the few GeV range, and the study of the internal
structure of prominent resonances in meson electroproduction.  

\section{Establishing the N* Spectrum}
\label{sec:1}
The complex structure of the light-quark (u \& d quarks) baryon excitation spectrum complicates the experimental 
search for individual states. As a result of the strong interaction, resonances are wide, often 200 MeV to 400 MeV, 
and are difficult to uniquely identify when only differential cross sections are measured. Most of the excited 
nucleon states listed in the Review 
of Particle Properties prior to 2012 have been observed in elastic pion scattering 
$\pi N \to \pi N$. However there are important limitations in the sensitivity to the 
higher mass nucleon states that may have small $\Gamma_{\pi N}$ decay widths.  
The extraction of resonance contributions then becomes exceedingly difficult in $\pi N $ scattering. 
Estimates for alternative decay channels have 
been made in quark model calculations\cite{Capstick:1993kb} for various channels.  This has
 led to a major experimental effort at Jefferson Lab, ELSA, GRAAL, and MAMI
to chart differential cross sections and polarization observables for a variety of meson
 photoproduction channels. At JLab with CLAS, several final states have 
 been measured with high precision\cite{Dugger:2005my,Dugger:2009pn,Williams:2009yj,
 Williams:2009aa,Williams:2009ab,Bradford:2006ba,Bradford:2005pt,McCracken:2009ra,
 Dey:2010hh,McNabb:2003nf} that are now employed in multi-channel analyses.  
\begin{figure}[h]
\centering
\resizebox{0.65\columnwidth}{!}{\includegraphics{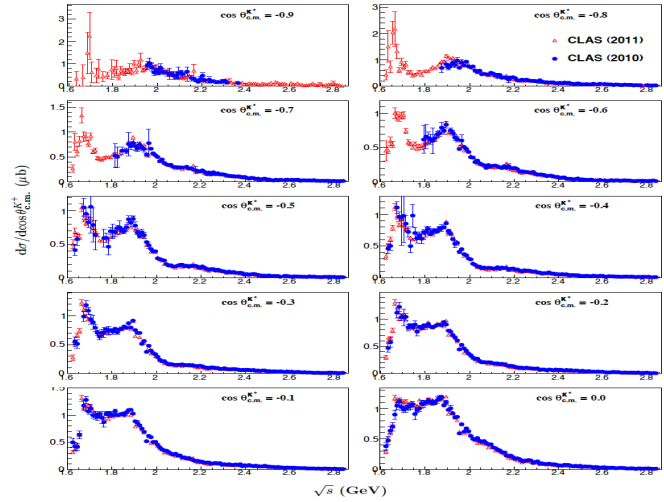}}
\caption{Invariant mass dependence of the $\gamma p \to K^+\Lambda$ differential cross section in the 
backward polar angle range. There are 3 structure visible that may indicate resonance excitations, 
at 1.7, 1.9, and 2.2 GeV. The blue full circles are based on the topology $K^+p\pi^-$, 
the red open triangles
are based on topology $K^+p$ or $K^+\pi^-$, which extended coverage towards lower W at backward angles
and allows better access to the resonant structure near threshold. }
\label{KLambda-crs}
\end{figure}

\subsection{New states from open strangeness photoproduction}
 \label{KLambda}
 Here one focus has recently been on measurements of $\gamma p \to K^+\Lambda$, using a 
 polarized photon beam several polarization observables can be measured by analyzing the 
 parity violating decay of the recoil $\Lambda \to p \pi^-$. It is well known that the energy-dependence of 
 a partial-wave amplitude for one particular channel is influenced by other reaction 
channels due to unitarity constraints. To fully describe the energy-dependence 
of an amplitude one has to include other reaction channels in a coupled-channel approach. 
Such analyses have been developed by the Bonn-Gatchina group\cite{Anisovich:2011fc}, 
at JLab\cite{JuliaDiaz:2007kz}, at J\"ulich\cite{Ronchen:2014cna} and other groups.

\begin{figure}[h]
\centering
\resizebox{0.65\columnwidth}{!}{\includegraphics{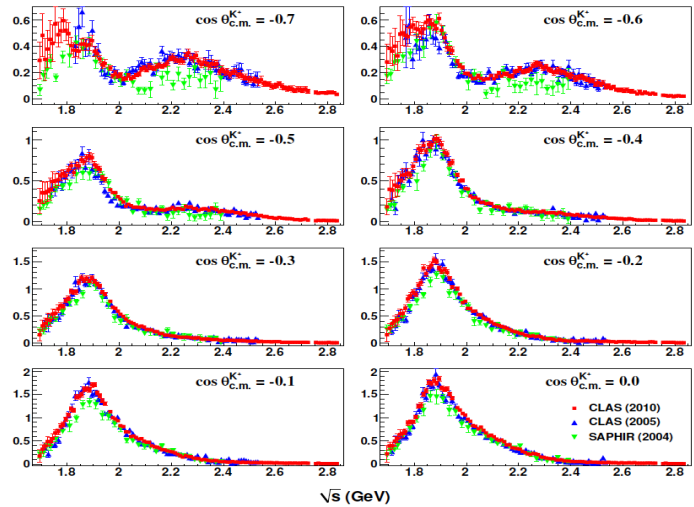}}
\caption{Invariant mass dependence of the $\gamma p \to K^+\Sigma^\circ$ differential cross section in the backward
polar angle range. }
\label{KSigma-crs}
\end{figure}

The data sets with the highest impact on resonance amplitudes in the mass range above 1.7~GeV have been 
kaon-hyperon production using a spin-polarized photon beam and 
where the polarization of the $\Lambda$ or $\Sigma^\circ$ is also measured. The high precision cross section and 
polarization data~\cite{Bradford:2006ba,Bradford:2005pt,McCracken:2009ra,Dey:2010hh,McNabb:2003nf} provide 
nearly full polar angle coverage and span the $K^+\Lambda$ invariant mass range 
from threshold to 2.9 GeV, hence covering the full nucleon resonance domain where new states might be discovered. 

The backward angle $K^+\Lambda$ data in Fig.\ref{KLambda-crs} show clear resonance-like structures 
at 1.7 GeV and 1.9 GeV that are particularly prominent and well-separated from other structures at backward angles, while 
at more forward angles (not shown) t-channel processes become prominent and dominate the cross section.
The broad enhancement at 2.2~GeV may also indicate resonant behavior although it is less visible at more 
central angles with larger background contributions. 
The $K^+\Sigma$ channel also indicates significant resonant behavior as seen in Fig.~\ref{KSigma-crs}. The peak structure at 1.9 GeV is 
present at all angles with a maximum strength near 90 degrees, consistent with the behavior of a $J^P= {3\over 2}^+$ 
p-wave. Other structures near 2.2 to 2.3~GeV are also visible. 
Still, only a full partial wave analysis can determine the underlying resonances, their masses and spin-parity.  
The task is somewhat easier for the $K\Lambda$ channel, as the iso-scalar nature of the $\Lambda$ selects 
isospin-${1\over 2}$ states to contribute to the $K\Lambda$ 
final state, while both isospin-${1\over 2}$ and isospin-${3\over 2}$ states can contribute to the $K\Sigma$ final state.     
\begin{figure}[t]
\centering
\resizebox{0.65\columnwidth}{!}{\includegraphics{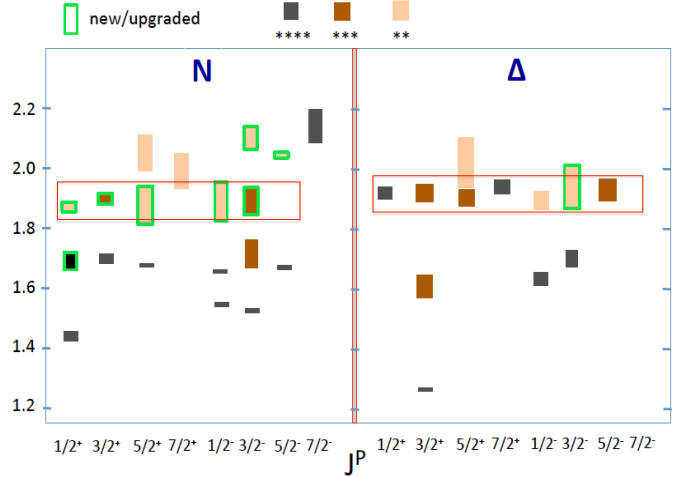}}
\caption{Nucleon and $\Delta$ resonance spectrum below 2.2 GeV in RPP 2014~\cite{Agashe:2014kda}. The
new states and states with improved evidence observed in the recent Bonn-Gatchina multi-channel analysis are 
shown with the green frame. The red frames highlight the apparent mass degeneracy of five or six states with 
different spin and parity.  The analysis includes all the $K^+\Lambda$ and $K^+\Sigma^\circ$ cross section and 
polarization data.}
\label{pdg2014}
\end{figure}
\begin{figure}[h]
\centering
\resizebox{0.50\columnwidth}{!}{\includegraphics{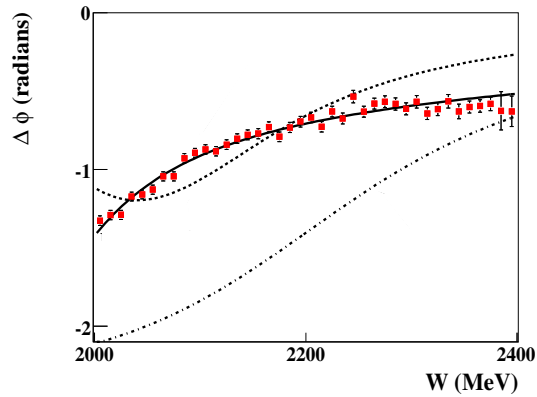}}
\caption{Phase motion of the partial wave fit to the $\gamma p \to p \omega$ differential cross section and spin density matrix elements. 3 resonant 
states, the subthreshold resonance $N(1680){5\over 2}^+$, $N(2190){7\over 2}^-$, and the missing $N(2000){5\over 2}^+$ are needed to fit the data (solid line). Fits without $N(2000){5\over 2}^+$ (dashed-dotted line), or without $N(1680){5\over 2}^+$ (dashed line) cannot reproduce the data.}
\label{omega}
\end{figure}

These cross section data together with the $\Lambda$ and $\Sigma$ recoil polarization and polarization transfer data 
to the $\Lambda$ and $\Sigma$ had strong impact on the discovery of several new nucleon 
states. They also provided new evidence for several candidate states that had been observed previously but lacked 
confirmation as shown in Fig.~\ref{pdg2014}. It is interesting to observe that five of the observed nucleon states have nearly 
degenerate masses near 1.9~GeV. 
Similarly, the new $\Delta$ state appears to complete a mass degenerate multiplet near 1.9~GeV as well. There is no obvious 
mechanism for this apparent degeneracy.  Nonetheless, all new states may be accommodated within the symmetric 
constituent quark model based on $SU(6)\otimes O(3)$ symmetry group as far as quantum numbers are concerned. As discussed 
in section~\ref{intro} for the case of the Roper resonance $N(1440){1\over 2}^+$, the masses of all pure  quark model states 
need to be corrected for dynamical coupled channel effects to compare them with observed resonances.  The same applies to 
the recent Lattice QCD predictions~\cite{Edwards:2011jj} for the nucleon and Delta spectrum. 

\begin{figure}[t]
\centering
\resizebox{0.65\columnwidth}{!}{\includegraphics{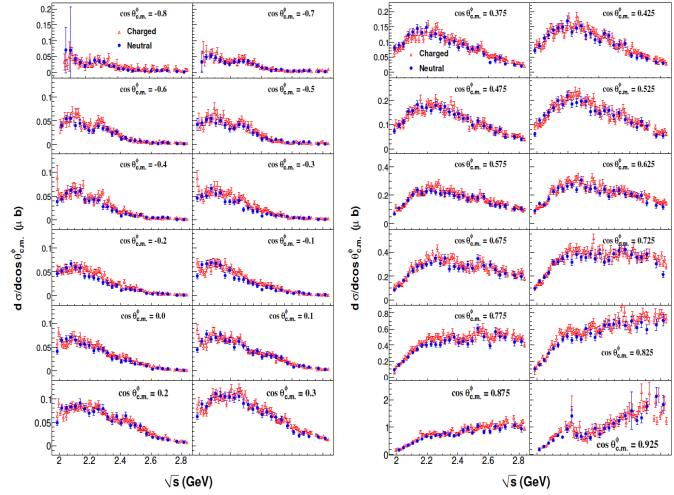}}
\caption{Differential cross sections in a nearly full angular range for $\gamma p \to p \phi$  production. }
\label{phi}
\end{figure}
\begin{figure}[h]
\vspace{-1cm}
\centering
\resizebox{0.6\columnwidth}{!}{\includegraphics{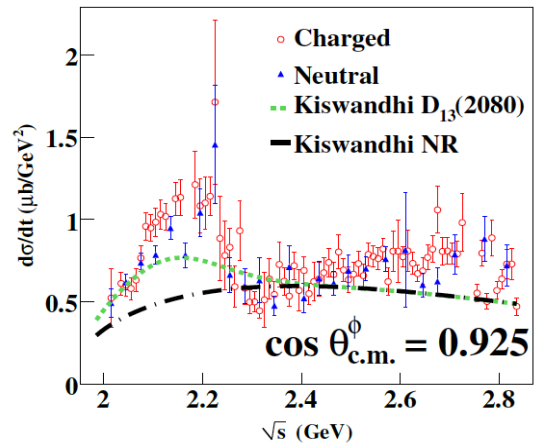}}
\caption{Differential cross sections of $\gamma p \to p \phi$  production for the most forward angle bin. The two curves 
refer to fits without (dashed) and with (dotted) a known resonance at 2.08 GeV included. }
\label{phi_forward}
\end{figure}
\vspace{-1cm}
\subsection{Vectormeson photoproduction}
\label{Vectormeson}

In the mass range above 2.0~GeV resonances tend to decouple from simple final states like $N\pi$, $N\eta$, and 
$K\Lambda$. We have to consider more complex final states with multi-mesons or vector mesons, such 
as $N\omega$, $N\phi$, and $K^*\Sigma$. The study of such final states adds significant complexity as more amplitudes can 
contribute to spin-1 mesons photoproduction compared to pseudo-scalar meson production. As is the case for $N\eta$ production, 
the $N\omega$ channel is selective to isospin $1\over 2$ nucleon states only.       
CLAS has collected a tremendous amount of data in the $p\omega$~\cite{Williams:2009aa,Williams:2009ab} and 
$p\phi$~\cite{Seraydaryan:2013ija,Dey:2014tfa} final states on differential cross sections and spin-density matrix elements that are now entering 
into the more complex multi-channel analyses such as Bonn-Gatchina~\cite{sarantsev2015}.  The CLAS collaboration performed a single channel event-based analysis, whose results are shown in Fig.~\ref{omega}, and provide further evidence for the $N(2000){5\over 2}^+$.  

Photoproduction of $\phi$ mesons is also considered a potentially rich source of new excited nucleon states in the mass range above 2 GeV. 
Some lower mass states such as $N(1535){1\over 2}^-$ may have significant $s\bar{s}$ components~\cite{Liu:2005pm}. Such components 
may result in states coupling to $p\phi$ with significant strength above threshold. Differential cross sections and spin-density matrix elements 
have been measured for $\gamma p \to p \phi$ in a mass range up to nearly 3 GeV. In Fig.~\ref{phi} structures are seen near 2.2~GeV in the forward most angle bins and at very backward angles for both decay channels $\phi \to K^+K^-$ and 
$\phi \to K_l^0K_s^0$, and with the exception of the smallest forward angle bin the structures are more prominent at backward angles. 
Only a multi-channel partial wave analysis will be able to pull out any significant resonance strength.  
Fig.~\ref{phi_forward} shows the differential cross section 
$d\sigma /dt$ 
of the most forward angle bin. A broad structure at 2.2 GeV is present, but does not show the typical Breit-Wigner behavior of a single 
resonance. It also does 
not fit the data in a larger angle range, which indicates that contributions other than genuine resonances may be significant. 
The forward and backward angle structures 
may also hint at the presence of dynamical effects possibly due to molecular contributions such as diquark-anti-triquark 
contributions~\cite{Lebed:2015dca}, the strangeness equivalent to the recently observed hidden charm $P_c^+$ states.   
    
Another process that has promise in the search for new excited baryon states, including those with 
isospin-${3\over 3}$ is $\gamma p \to K^*\Sigma$~\cite{sarantsev2015}. 
In distinction to the vector mesons discussed above, diffractive processes do not play a role in this channel, which then 
should allow better direct access to s-channel resonance production.    
\begin{figure}[b]
\vspace{-1.0cm}
\centering
\resizebox{0.65\columnwidth}{!}{\includegraphics{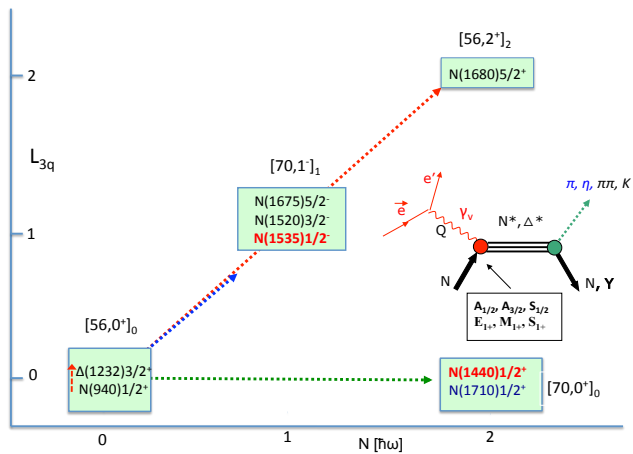}}
\vspace{-0.75cm}
\caption{Schematic of $SU(6)\otimes O(3)$ supermultiplets with excited states that have been explored in $e p \to e^\prime \pi^+ n$, 
$e p \to e^\prime p^\prime \pi^\circ$ and $ep \to e^\prime p^\prime \pi^+\pi^-$. The insert shows the helicity amplitudes and electromagnetic multipoles  extracted from the data. Only the ones highlighted in red are discussed here.}
\label{su6}
\end{figure}
\section{Structure of excited nucleons}
\label{structure}
\begin{figure}[t]
\vspace{-0.5cm}
\centering
\resizebox{0.70\columnwidth}{!}{\includegraphics{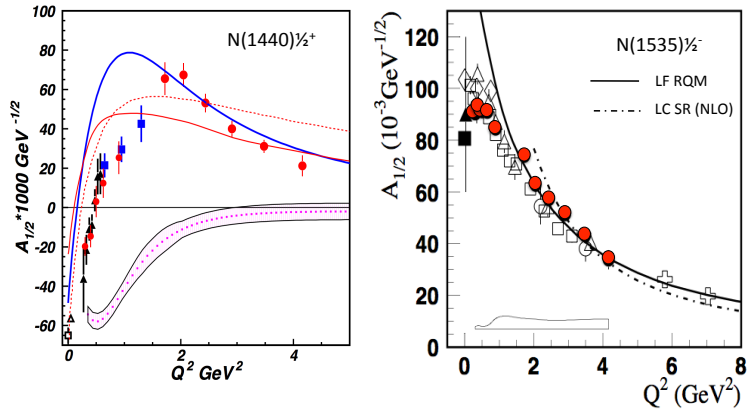}}
\vspace{-1.5cm}
\caption{Left panel: The transverse helicity amplitudes $A_{1/2}$ for the Roper resonance $N(1440){1\over 2}^+$. Data are from CLAS compared
to two LF RQM with fixed quark masses (dashed) and with running quark mass (solid red), and with projections from the DSE/QCD approach. The 
magenta dotted line with error band indicates non 3-quark contributions obtained from a the difference of the DSE curve and the CLAS data. The right panel shows the same amplitude for the $N(1535){1\over 2}^-$ compared to LF RQM calculations (solid line) 
and QCD computation within the LC Sum Rule approach.}
\label{p11_s11}
\end{figure}

Meson photoproduction has become an essential tool in the search for new excited baryons. The exploration of the internal structure 
of excited states and the effective degrees of freedom contributing to s-channel resonance excitation requires use of electron beams 
where the virtuality ($Q^2$) of the exchanged photon can be varied to probe the spatial structure (Fig.~\ref{su6}). 
Electroproduction of final states with pseudoscalar mesons 
(e.g. $N\pi$, $p\eta$, $K\Lambda$) have been employed with CLAS, leading to new insights into the scale dependence of 
effective degrees of freedom, e.g. meson-baryon, constituent quark, and dressed quark contributions. Several excited states, shown 
in Fig.~\ref{su6} assigned to their primary $SU(6) \otimes O(3)$ supermultiplets have been studied. The $N\Delta(1232){3\over 2}^+$ transition is now well measured in a large range of $Q^2$~\cite{Joo:2001tw,Ungaro:2006df,Frolov:1998pw,Aznauryan:2009mx}. 
Two of the prominent higher mass states, the Roper resonance
$N(1440){1\over 2}^+$  and $N(1535){1\over 2}^-$ are shown in Fig.~\ref{p11_s11} as representative 
examples~\cite{Aznauryan:2008pe,Aznauryan:2009mx} from a wide program at 
JLab~\cite{Mokeev:2012vsa,Denizli:2007tq,Armstrong:1998wg,Egiyan:2006ks,Park:2007tn,Park:2014yea}.  For these two states advanced
relativistic quark model calculations~\cite{Aznauryan:2015zta} and QCD calculations from Dyson-Schwinger Equation~\cite{Segovia:2015hra} and 
Light Cone sum rule~\cite{Anikin:2015ita} have recently become available, for the first time employing QCD-based modeling of the excitation 
of the quark core. There is agreement with the data at $Q^2 > 1.5$~GeV$^2$. The calculations deviate significantly from the data 
at lower $Q^2$, which indicates significant  non quark core effects.  For the Roper resonance such contributions have been described 
successfully in dynamical meson-baryon models~\cite{Obukhovsky:2011sc} and in effective field theory~\cite{Bauer:2014cqa}. 
\begin{figure}[h]
\vspace{-0.4cm}
\centering
\resizebox{0.75\columnwidth}{!}{\includegraphics{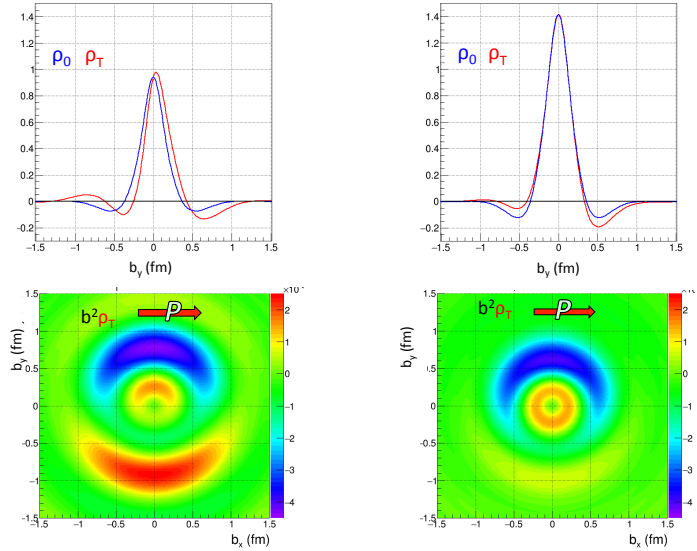}}
\vspace{-0.5cm}
\caption{Left panels: $N(1440)$, top: projection of charge densities on $b_y$, bottom: transition charge densities when the proton  
is spin polarized along $b_x$. Right panels: same for $N(1535)$. Note that the densities are scaled with 
$b^2$ to emphasize the outer wings. Color code:negative charge is blue, positive charge is red. Note that all scales
are the same. }
\label{charge_densities}
\end{figure}

Knowledge of the helicity amplitudes in a large $Q^2$ allows for the determination of the transition charge densities on the light 
cone in transverse impact parameter space ($b_x, b_y$)~\cite{Tiator:2008kd}. Figure~\ref{charge_densities} shows the comparison 
of $N(1440){1\over 2}^+$ and  $N(1535){1\over 2}^-$. There are clear differences in the charge transition densities between 
the two states. The Roper state has a softer positive core and a wider negative outer cloud than $N(1535)$ 
and develops a larger shift in $b_y$ when the proton is polarized along the $b_x$ axis.     

\section{Conclusions and Outlook}
Over the past five years eight baryon states in the mass range from 1.85 to 2.15 GeV have been either discovered or 
evidence for the existence of states has been significantly strengthened. To a large degree this is the result of adding 
very precise photoproduction data in open strangeness channels to the data base that is included in multi-channel partial wave 
analyses, especially the Bonn-Gatchina PWA. The possibility to measure polarization observables in these processes has been 
critical~\cite{pasyuk2015}. In the mass range above 2 GeV more complex processes such as vector mesons or $\Delta\pi$ 
may have sensitivity 
to states with higher masses but require more complex analyses techniques to be brought to bear. Precision data in such 
channels have been available for a few years but remain to be fully incorporated in multi-channel partial wave analyses processes. 
The light-quark baryon spectrum is likely also populated with hybrid excitations~\cite{Dudek:2012ag} where the gluonic 
admixtures to the wave function are dominating the excitation. These states appear with the same quantum numbers as 
ordinary quark excitations, and can only be isolated from ordinary states due to the $Q^2$ dependence of their helicity 
amplitudes~\cite{Li:1991yba}, which is expected to be quite different from ordinary quark excitations. This requires new 
electroproduction data especially at low $Q^2$~\cite{LOI_Hybrids} with different final states and at masses above 2 GeV.  
 
Despite the very significant progress made in recent years to further establish the light-quark baryon spectrum and explore 
the internal structure of excited states, much remains to be done. A vast amount of precision data already collected 
needs  to be included in the multi-channel analysis frameworks, and polarization data are still to be analyzed~\cite{pasyuk2015}. 
There are approved 
proposals to study resonance excitations at much higher $Q^2$ and with higher precision at Jefferson Lab with CLAS12~\cite{Gothe} 
that may reveal the transition to the bare quark core contributions at short distances.   
%\vspace{-0.5cm}
\section{Acknowledgment}
I like to thank Inna Aznauryan and Viktor Mokeev for numerous discussions on the subjects discussed in this presentation. 
This work was supported by the US Department of Energy under contract No. DE-AC05-06OR23177.

%\vspace{-0.2cm}


\begin{thebibliography}{99}

\bibitem{Anderson:1952nw} 
  H.~L.~Anderson, E.~Fermi, E.~A.~Long and D.~E.~Nagle (1952)
  %``Total Cross-sections of Positive Pions in Hydrogen,''
  Phys.\ Rev.\  {\bf 85}, 936 . 
 
\bibitem{GellMann1964} 
  M. Gell-Mann, 
  Phys. Lett.  {\bf 8}, 214 (1964).

\bibitem{Zweig1964} 
  G. Zweig, CERN Reports, TH 401 and 412  (1964).

\bibitem{Greenberg:1964pe} 
  O.~W.~Greenberg,
  %``Spin and Unitary Spin Independence in a Paraquark Model of Baryons and Mesons,''
  Phys.\ Rev.\ Lett.\  {\bf 13}, 598 (1964); arXiv:0803.0992 [physics.hist-ph].

\bibitem{Burkert:2004sk} 
  V.~D.~Burkert and T.~S.~H.~Lee 
  %``Electromagnetic meson production in the nucleon resonance region,''
  Int.\ J.\ Mod.\ Phys.\ E {\bf 13}, 1035 (2004).

\bibitem{Agashe:2014kda} 
  K.~A.~Olive {\it et al.} [Particle Data Group],
  Chin.\ Phys.\ C {\bf 38}, 090001 (2014).

\bibitem{Dudek:2012ag} 
  J.~J.~Dudek and R.~G.~Edwards,
  Phys.\ Rev.\ D {\bf 85}, 054016 (2012)

\bibitem{Suzuki:2009nj} 
  N.~Suzuki et al., 
  %``Disentangling the Dynamical Origin of P-11 Nucleon Resonances,''
  Phys.\ Rev.\ Lett.\  {\bf 104}, 042302 (2010)
 % [arXiv:0909.1356 [nucl-th]].
  
\bibitem{Klempt:2009pi} 
  E.~Klempt and J.~M.~Richard,
  %``Baryon spectroscopy,''
  Rev.\ Mod.\ Phys.\  {\bf 82}, 1095 (2010)

\bibitem{Tiator:2011pw} 
  L.~Tiator, D.~Drechsel, S.~S.~Kamalov and M.~Vanderhaeghen,
  %``Electromagnetic Excitation of Nucleon Resonances,''
  Eur.\ Phys.\ J.\ ST {\bf 198}, 141 (2011).   

\bibitem{Aznauryan:2011qj} 
  I.~G.~Aznauryan and V.~D.~Burkert,
  %``Electroexcitation of nucleon resonances,''
  Prog.\ Part.\ Nucl.\ Phys.\  {\bf 67}, 1 (2012)

\bibitem{Aznauryan:2012ba} 
  I.~G.~Aznauryan {\it et al.},
  %``Studies of Nucleon Resonance Structure in Exclusive Meson Electroproduction,''
  Int.\ J.\ Mod.\ Phys.\ E {\bf 22}, 1330015 (2013)

\bibitem{Crede:2013sze} 
  V.~Crede and W.~Roberts,
  %``Progress towards understanding baryon resonances,''
  Rept.\ Prog.\ Phys.\  {\bf 76}, 076301 (2013)

\bibitem{Bazavov:2014xya} 
  A.~Bazavov {\it et al.},
  %``Additional Strange Hadrons from QCD Thermodynamics and Strangeness Freezeout in Heavy Ion Collisions,''
  Phys.\ Rev.\ Lett.\  {\bf 113}, no. 7, 072001 (2014) [arXiv:1404.6511 [hep-lat]].

\bibitem{Bazavov:2014yba} 
  A.~Bazavov {\it et al.},
  %``The melting and abundance of open charm hadrons,''
  Phys.\ Lett.\ B {\bf 737}, 210 (2014) [arXiv:1404.4043 [hep-lat]].

\bibitem{Capstick:1993kb} 
  S.~Capstick and W.~Roberts,
  %``Quasi two-body decays of nonstrange baryons,''
  Phys.\ Rev.\ D {\bf 49}, 4570 (1994)
  
\bibitem{Dugger:2005my} 
  M.~Dugger {\it et al.},
  %``Eta-prime photoproduction on the proton for photon energies from 1.527-GeV to 2.227-GeV,''
  Phys.\ Rev.\ Lett.\  {\bf 96}, 062001 (2006)
  [Phys.\ Rev.\ Lett.\  {\bf 96}, 169905 (2006)]

\bibitem{Dugger:2009pn} 
  M.~Dugger {\it et al.} [CLAS Collaboration],
  %``pi+ photoproduction on the proton for photon energies from 0.725 to 2.875-GeV,''
  Phys.\ Rev.\ C {\bf 79}, 065206 (2009)
  doi:10.1103/PhysRevC.79.065206

\bibitem{Williams:2009yj} 
  M.~Williams {\it et al.} [CLAS Collaboration],
  %``Differential cross s ections for the reactions gamma p ---> p eta and gamma p ---> p eta-prime,''
  Phys.\ Rev.\ C {\bf 80}, 045213 (2009)

 \bibitem{Williams:2009aa} 
  M.~Williams {\it et al.} [CLAS Collaboration],
  %``Partial wave analysis of the reaction gamma p ---> p omega and the search for nucleon resonances,''
  Phys.\ Rev.\ C {\bf 80}, 065209 (2009)

\bibitem{Williams:2009ab} 
  M.~Williams {\it et al.} [CLAS Collaboration],
  %``Differential cross sections and spin density matrix elements for the reaction gamma p ---> p omega,''
  Phys.\ Rev.\ C {\bf 80}, 065208 (2009)
     
\bibitem{Bradford:2006ba} 
  R.~K.~Bradford {\it et al.} [CLAS Collaboration],
  %``First measurement of beam-recoil observables C(x) and C(z) in hyperon photoproduction,''
  Phys.\ Rev.\ C {\bf 75}, 035205 (2007)

\bibitem{Bradford:2005pt} 
  R.~Bradford {\it et al.} [CLAS Collaboration],
  %``Differential cross sections for gamma + p ---> K+ + Y for Lambda and Sigma0 hyperons,''
  Phys.\ Rev.\ C {\bf 73}, 035202 (2006)

\bibitem{McCracken:2009ra} 
  M.~E.~McCracken {\it et al.} [CLAS Collaboration],
  %``Differential cross section and recoil polarization measurements for the gamma p to K+ Lambda reaction using CLAS at Jefferson Lab,''
  Phys.\ Rev.\ C {\bf 81}, 025201 (2010)    

\bibitem{Dey:2010hh} 
  B.~Dey {\it et al.} [CLAS Collaboration],
  %``Differential cross sections and recoil polarizations for the reaction \gamma p -> K^+ \Sigma^0,''
  Phys.\ Rev.\ C {\bf 82}, 025202 (2010)
  
\bibitem{McNabb:2003nf} 
  J.~W.~C.~McNabb {\it et al.} [CLAS Collaboration],
  %``Hyperon photoproduction in the nucleon resonance region,''
  Phys.\ Rev.\ C {\bf 69}, 042201 (2004)
%%%%%%%%%%%%%%%%%%%%%%%%%%%%%%%%

\bibitem{Anisovich:2011fc} 
  A.~Anisovich, R.~Beck, E.~Klempt, V.~Nikonov, A.~Sarantsev and U.~Thoma,
  %``Properties of baryon resonances from a multichannel partial wave analysis,''
  Eur.\ Phys.\ J.\ A {\bf 48}, 15 (2012)
  
\bibitem{JuliaDiaz:2007kz} 
  B.~Julia-Diaz, T.-S.~H.~Lee, A.~Matsuyama and T.~Sato,
  %``Dynamical coupled-channel model of pi N scattering in the W <= 2-GeV nucleon resonance region,''
  Phys.\ Rev.\ C {\bf 76}, 065201 (2007)

\bibitem{Ronchen:2014cna} 
  D.~R\"onchen {\it et al.},
  %``Photocouplings at the Pole from Pion Photoproduction,''
  Eur.\ Phys.\ J.\ A {\bf 50}, no. 6, 101 (2014)

\bibitem{Edwards:2011jj} 
  R.~G.~Edwards, J.~J.~Dudek, D.~G.~Richards and S.~J.~Wallace,
  %``Excited state baryon spectroscopy from lattice QCD,''
  Phys.\ Rev.\ D {\bf 84}, 074508 (2011)

\bibitem{Seraydaryan:2013ija} 
  H.~Seraydaryan {\it et al.} [CLAS Collaboration],
  %``$\phi$-meson photoproduction on Hydrogen in the neutral decay mode,''
  Phys.\ Rev.\ C {\bf 89}, no. 5, 055206 (2014)

\bibitem{Dey:2014tfa} 
  B.~Dey {\it et al.} [CLAS Collaboration],
  %``Data analysis techniques, differential cross sections, and spin density matrix elements for the reaction $\gamma p \rightarrow \phi p$,''
  Phys.\ Rev.\ C {\bf 89}, no. 5, 055208 (2014)
 
\bibitem{sarantsev2015} A. Sarantsev, talk at this workshop  
 
\bibitem{Liu:2005pm} 
  B.~C.~Liu and B.~S.~Zou,
  %``Mass and K Lambda coupling of N*(1535),''
  Phys.\ Rev.\ Lett.\  {\bf 96}, 042002 (2006)    

\bibitem{Lebed:2015dca} 
  R.~F.~Lebed,
  %``Do the $P_c^+$ pentaquarks have strange siblings?,''
  Phys.\ Rev.\ D {\bf 92}, no. 11, 114030 (2015)

\bibitem{Joo:2001tw} 
  K.~Joo {\it et al.} [CLAS Collaboration],
  %``Q**2 dependence of quadrupole strength in the gamma* p ---> Delta+(1232) ---> p pi0 transition,''
  Phys.\ Rev.\ Lett.\  {\bf 88}, 122001 (2002)

\bibitem{Ungaro:2006df} 
  M.~Ungaro {\it et al.} [CLAS Collaboration],
  %``Measurement of the N ---> Delta+(1232) transition at high momentum transfer by pi0 electroproduction,''
  Phys.\ Rev.\ Lett.\  {\bf 97}, 112003 (2006)

\bibitem{Frolov:1998pw} 
  V.~V.~Frolov {\it et al.},
  %``Electroproduction of the Delta (1232) resonance at high momentum transfer,''
  Phys.\ Rev.\ Lett.\  {\bf 82}, 45 (1999)

\bibitem{Aznauryan:2008pe} 
  I.~G.~Aznauryan {\it et al.} [CLAS Collaboration],
  %``Electroexcitation of the Roper resonance for 1.7 < Q**2 < 4.5 -GeV2 in vec-ep ---> en pi+,''
  Phys.\ Rev.\ C {\bf 78}, 045209 (2008)

 \bibitem{Aznauryan:2009mx} 
  I.~G.~Aznauryan {\it et al.} [CLAS Collaboration], 
  %``Electroexcitation of nucleon resonances from CLAS data on single pion electroproduction,''
  Phys.\ Rev.\ C {\bf 80}, 055203 (2009)
 
\bibitem{Mokeev:2012vsa} 
  V.~I.~Mokeev {\it et al.} [CLAS Collaboration],
  %``Experimental Study of the $P_{11}(1440)$ and $D_{13}(1520)$ resonances from CLAS data on $ep \rightarrow e'\pi^{+} \pi^{-} p'$,''
  Phys.\ Rev.\ C {\bf 86}, 035203 (2012)

\bibitem{Denizli:2007tq} 
  H.~Denizli {\it et al.} [CLAS Collaboration],
  %``Q*2 dependence of the S(11)(1535) photocoupling and evidence for a P-wave resonance in eta electroproduction,''
  Phys.\ Rev.\ C {\bf 76}, 015204 (2007)

\bibitem{Armstrong:1998wg} 
  C.~S.~Armstrong {\it et al.} [Jefferson Lab E94014 Collaboration],
  %``Electroproduction of the S(11)(1535) resonance at high momentum transfer,''
  Phys.\ Rev.\ D {\bf 60}, 052004 (1999)
   
 \bibitem{Egiyan:2006ks} 
  H.~Egiyan {\it et al.} [CLAS Collaboration],
  %``Single pi+ electroproduction on the proton in the first and second resonance regions at 0.25-GeV**2 < Q**2 < 0.65-GeV**2 using CLAS,''
  Phys.\ Rev.\ C {\bf 73}, 025204 (2006) 
  
\bibitem{Park:2007tn} 
  K.~Park {\it et al.} [CLAS Collaboration],
  %``Cross sections and beam asymmetries for vec(e) p ---> en pi+ in the nucleon resonance region for 1.7 <= Q**2 <= 4.5-(GeV)**2,''
  Phys.\ Rev.\ C {\bf 77}, 015208 (2008)

\bibitem{Park:2014yea} 
  K.~Park {\it et al.} [CLAS Collaboration],
  %``Measurements of $ep \to e^\prime \pi^+n$ at W = 1.6 - 2.0 GeV and extraction of nucleon resonance electrocouplings at CLAS,''
  Phys.\ Rev.\ C {\bf 91}, 045203 (2015)

\bibitem{Aznauryan:2015zta} 
  I.~G.~Aznauryan and V.~D.~Burkert,
  %``Electroexcitation of the $\Delta(1232)\frac{3}{2}^+$ and $\Delta(1600)\frac{3}{2}^+$ in a light-front relativistic quark model,''
  Phys.\ Rev.\ C {\bf 92}, no. 3, 035211 (2015)

\bibitem{Segovia:2015hra} 
  J.~Segovia et al.,   %``Completing the picture of the Roper resonance,''
  Phys.\ Rev.\ Lett.\  {\bf 115}, no. 17, 171801 (2015)

\bibitem{Anikin:2015ita} 
  I.~V.~Anikin, V.~M.~Braun and N.~Offen,
  %``Electroproduction of the $N^*(1535)$ nucleon resonance in QCD,''
  Phys.\ Rev.\ D {\bf 92}, no. 1, 014018 (2015)

\bibitem{Obukhovsky:2011sc} 
  I.~T.~Obukhovsky et al., 
  %``Electroproduction of the Roper resonance on the proton: the role of the three-quark core and the molecular $N\sigma$ component,''
  Phys.\ Rev.\ D {\bf 84}, 014004 (2011)
  
\bibitem{Bauer:2014cqa} 
  T.~Bauer, S.~Scherer and L.~Tiator,
  %``Electromagnetic transition form factors of the Roper resonance in effective field theory,''
  Phys.\ Rev.\ C {\bf 90}, no. 1, 015201 (2014)

\bibitem{Tiator:2008kd} 
  L.~Tiator and M.~Vanderhaeghen,
  %``Empirical transverse charge densities in the nucleon-to-P(11)(1440) transition,''
  Phys.\ Lett.\ B {\bf 672}, 344 (2009)

 \bibitem{pasyuk2015} E. Pasyuk, talk at this workshop. 
 
\bibitem{Li:1991yba} 
  Z.~p.~Li, V.~Burkert and Z.~j.~Li,
  %``Electroproduction of the Roper resonance as a hybrid state,''
  Phys.\ Rev.\ D {\bf 46}, 70 (1992).

\bibitem{LOI_Hybrids}  
 L. Lanza, talk at this workshop;  A. D' Angelo et al., Jefferson Lab Letter of Intent LOI12-15-004 (2015). 

 
\bibitem{Gothe} R. Gothe, D. Carman,  V. Mokeev,  K. Park, talks at this workshop.
 
\end{thebibliography}
\end{document}